\documentclass[prl,twocolumn,showpacs,preprintnumbers,amsmath,amssymb]{revtex4}
\usepackage{graphicx}
\usepackage{dcolumn}
\usepackage{bm}
\usepackage{amsthm}
\usepackage{color}
\input xy
\xyoption{all}

\newcommand{\ket}[1]{\left| #1 \right\rangle}
\newcommand{\bra}[1]{\left\langle #1 \right|}

\newcommand{\twotwo}[4]{\left[ \begin{array}{cc} #1 & #2 \\
                        #3 & #4 \end{array} \right]}

\DeclareMathOperator{\tr}{tr}
\DeclareMathOperator{\PTr}{PTr}
\DeclareMathOperator{\ptr}{ptr}

\begin{document}

\title{Only $n$-Qubit Greenberger-Horne-Zeilinger States
are Undetermined by their Reduced Density Matrices}

\author{Scott N. Walck}
  \email{walck@lvc.edu}
\author{David W. Lyons}
  \email{lyons@lvc.edu}
\affiliation{Lebanon Valley College, Annville, PA 17003}

\date{December 4, 2007}

\begin{abstract}
The generalized $n$-qubit Greenberger-Horne-Zeilinger (GHZ) states
and their local unitary equivalents
are the only states of $n$ qubits that are not uniquely determined
among pure states by their reduced density matrices of $n-1$ qubits.
Thus, among pure states, the generalized GHZ states are the only ones
containing information at the $n$-party level.
We point out a connection between
local unitary stabilizer subgroups and the property of
being determined by reduced density matrices.
\end{abstract}


\maketitle

Quantifying and characterizing multi-party
quantum entanglement is a fundamental
problem in the field of quantum information.
Roughly speaking, one expects the states that are ``most entangled''
to be the most valuable resources for carrying out quantum
information processing tasks such as quantum communication
and quantum teleportation, and to give the most striking
philosophical implications in terms of the rejection
of local hidden variable theories \cite{bouwmeester99}.

Although no single definition of ``most entangled'' seems possible,
since we know that multi-party
entanglement occurs in many types that admit at best
a partial order \cite{dur00},
it is
still worthwhile to consider properties that carry some of the spirit
of ``most entangled.''

One such property is the failure of a state to be determined
by its reduced density matrices.
As reduced density matrices contain correlation information pertaining
to fewer than the full number of parties in the system,
states exhibiting entanglement involving \emph{all} parties must
possess information beyond that contained in their reduced density matrices.
Linden, Popescu, and Wootters put forward this suggestion
in \cite{linden02,linden02b} and proved
the surprising result that almost all $n$-party pure states
\emph{are determined} by their reduced density matrices.
In other words, the set of $n$-party pure states undetermined
by their reduced density matrices is a set of measure zero.
In \cite{diosi04}, Di\'osi gave a constructive method
that succeeds in almost all cases
for determining a 3-qubit pure state from its reduced
density matrices.
Nevertheless, the question of precisely which states
are determined by their reduced density matrices
remained open.

In this Letter, we show that the only $n$-qubit states
that are undetermined among pure states by their reduced
density matrices are the generalized $n$-qubit
Greenberger-Horne-Zeilinger (GHZ) states,
\[
\alpha \ket{0 0 \cdots 0} + \beta \ket{1 1 \cdots 1} , \hspace{5mm}
 \alpha, \beta \neq 0
\]
and their local unitary (LU) equivalents.
This means that, among pure states,
the generalized GHZ states are the only ones
containing information at the $n$-party level.
For the case $n=3$, this result was reported previously
in \cite{linden02}.
Part of our argument employs the methods
of \cite{diosi04} in an essential way.

Let $D_n$ be the set of $n$-qubit density matrices.
If $\rho \in D_n$ is an $n$-qubit density matrix,
and $j \in \{1,\ldots,n\}$ is a qubit label, we may
form an $(n-1)$-qubit reduced density matrix
$\rho_{(j)} = \tr_j \rho$
by taking the partial trace of $\rho$ over qubit $j$.
Let
\[
\PTr: D_n \to D_{n-1}^n
\]
be the map
$\rho \mapsto (\rho_{(1)},\ldots,\rho_{(n)})$
that associates to $\rho$ its $n$-tuple
of $(n-1)$-qubit reduced density matrices.
The map $\PTr$ is neither injective (one-to-one)
nor surjective (onto).
Its failure to be surjective means that there are
$n$-tuples of $(n-1)$-qubit density matrices that
cannot be produced from any $n$-qubit density matrix
by the partial trace.
The question of whether a collection of $(n-1)$-qubit
reduced density matrices could have come from an
$n$-qubit density matrix by the partial trace is
the subject of recent and ongoing investigations
\cite{han04,cai07}.
The failure of $\PTr$ to be injective means that
multiple $n$-qubit states can have the same reduced
density matrices.  States $\rho_1 \neq \rho_2$
with $\PTr(\rho_1) = \PTr(\rho_2)$
require more information for their determination
than is contained in their $(n-1)$-qubit reduced
density matrices.

Let
\[
P_n = \{ \rho \in D_n | \rho^2 = \rho \}
\]
be the set of pure $n$-qubit states.
If we are interested primarily in pure states,
we can restrict the partial trace map
to pure $n$-qubit states.
\[
\ptr = \PTr|_{P_n} : P_n \to D_{n-1}^n
\]
Given a pure state $\psi$ with $\ket{\psi} \bra{\psi} \in P_n$,
the set $\ptr^{-1}(\ptr(\psi))$ contains all pure states
with the same reduced density matrices as $\psi$.
(We abbreviate $\ptr(\psi) = \ptr(\ket{\psi} \bra{\psi})$.)

We define a state $\psi$ to be
\emph{determined among pure states}
if $\ptr^{-1}(\ptr(\psi))$ contains only $\ket{\psi} \bra{\psi}$,
and \emph{undetermined among pure states}
if $\ptr^{-1}(\ptr(\psi))$ contains more than one state.
Similarly, we define a state $\rho \in D_n$ to be
\emph{determined among arbitrary states}
if $\PTr^{-1}(\PTr(\rho))$ contains only $\rho$,
and \emph{undetermined among arbitrary states}
if $\PTr^{-1}(\PTr(\rho))$ contains more than one state.

The surprising result of Linden and Wootters \cite{linden02b}
is that almost all $n$-qubit pure states
are determined among arbitrary states.

Nevertheless, there are pure states
that are undetermined among pure states
(and consequently undetermined among arbitrary states).
For example, consider the one-parameter family
of $n$-qubit states
\[
\ket{\eta} = \frac{1}{\sqrt{2}} \ket{0 0 \cdots 0}
           + \frac{\eta}{\sqrt{2}} \ket{1 1 \cdots 1} ,
\]
where $\eta$ is a complex number with magnitude one.
If $\eta_1 \neq \eta_2$, then $\ket{\eta_1}$
and $\ket{\eta_2}$ are different states
with different density matrices
$\ket{\eta_1} \bra{\eta_1} \neq
\ket{\eta_2} \bra{\eta_2}$,
yet they share the same reduced density matrices,
that is,
$\ptr(\ket{\eta_1} \bra{\eta_1}) =
 \ptr(\ket{\eta_2} \bra{\eta_2})$.

We see that almost all pure $n$-qubit states
are determined among pure states,
yet $n$-qubit GHZ states
are undetermined among pure states.
The question then becomes, precisely which states $\psi$
are undetermined among pure states?

\paragraph{Main Result.}
An $n$-qubit state $\psi$ is
undetermined among pure states
if and only if
$\psi$ is LU-equivalent to a generalized $n$-qubit GHZ state.

\begin{proof}
Let $\psi$ be an $n$-qubit pure state.

Suppose that 
$\psi$ is LU-equivalent to a generalized $n$-qubit GHZ state,
so we have
\[
U \ket{\psi} = \alpha \ket{0 0 \cdots 0}
             + \beta \ket{1 1 \cdots 1} ,
\]
where $U$ is a local unitary transformation.
Define 
$U \ket{\psi'} = \alpha \ket{0 0 \cdots 0}
               - \beta \ket{1 1 \cdots 1}$.
Then $\ket{\psi} \bra{\psi} \neq \ket{\psi'} \bra{\psi'}$,
since $\alpha \beta \neq 0$, but $\ptr(\psi) = \ptr(\psi')$.
Hence, $\psi$ is undetermined among pure states.

Conversely,
suppose that $\ket{\psi}$ is
undetermined among pure states.
Then there is an $n$-qubit state vector
$\ket{\psi'} \neq e^{i \alpha} \ket{\psi}$ that has the same
reduced density matrices as $\ket{\psi}$.

Claim:  If $\ket{\psi}$ and $\ket{\psi'}$ have the same reduced
density matrices, then for each qubit $j \in \{1,\ldots,n\}$,
there is a one-qubit local unitary transformation $L_j$ such that
$\ket{\psi'} = L_j \ket{\psi}$.

To prove this,
let $j \in \{1,\ldots,n\}$ be a qubit label.
Let $\rho_j$ denote the one-qubit reduced density matrix
of $\ket{\psi}$ for qubit $j$.
We write $\rho_j$ as a spectral decomposition,
\[
\rho_j = \sum_{i_j=0}^1 p_j^{i_j} \ket{i_j} \bra{i_j} ,
\]
for some orthonormal basis $\ket{i_j}$, where
$p_j^0$ and $p_j^1$ are the eigenvalues of $\rho_j$.
If $p_j^0 \neq p_j^1$, then
the orthonormal basis $\ket{i_j}$ is uniquely determined
up to a phase.
If $p_j^0 = p_j^1$, then any one-qubit orthonormal basis
can be used.

The $(n-1)$-qubit reduced density matrix
\[
\rho_{(j)} = \tr_j \ket{\psi} \bra{\psi} ,
\]
obtained by taking the
partial trace of $\ket{\psi} \bra{\psi}$ over qubit $j$,
has the same nonzero eigenvalues
as $\rho_j$,
\[
\rho_{(j)} = \sum_{i_j=0}^1 p_j^{i_j} \ket{i_j;(j)} \bra{i_j;(j)} .
\]
If $p_j^0 \neq p_j^1$, then the $(n-1)$-qubit
eigenvectors $\ket{0;(j)}$ and $\ket{1;(j)}$
are unique up to a phase.
If $p_j^0 = p_j^1$, then the eigenvectors of $\rho_{(j)}$
with eigenvalue
$p_j^0 = p_j^1 = 1/2$ constitute a two-dimensional subspace
of the $2^{n-1}$-dimensional vector space of $(n-1)$-qubit vectors,
and any orthonormal pair of vectors in this subspace may be chosen
as a basis.

We choose the one-qubit orthonormal basis $\ket{i_j}$
and the $(n-1)$-qubit orthonormal basis $\ket{i_j;(j)}$
so that $\ket{\psi}$ can be written
\[
\ket{\psi} = \sqrt{p_j^0} \ket{0} \otimes_j \ket{0;(j)}
           + \sqrt{p_j^1} \ket{1} \otimes_j \ket{1;(j)} ,
\]
where $\otimes_j$ is the tensor product that inserts a one-qubit ket
just before the $j$th factor in the $(n-1)$-qubit ket $\ket{i_j;(j)}$.

Now $\ket{\psi'}$ can be regarded as the state of a bipartite system
composed of qubit $j$ and all qubits but $j$, and has a Schmidt
decomposition with respect to those subsystems,
\[
\ket{\psi'} = \sqrt{q_j^0} \ket{0'} \otimes_j \ket{0';(j)}
            + \sqrt{q_j^1} \ket{1'} \otimes_j \ket{1';(j)} ,
\]
where $\ket{0'},\ket{1'}$ are orthonormal one-qubit vectors and
$\ket{0';(j)},\ket{1';(j)}$ are orthonormal $(n-1)$-qubit vectors.
Taking the partial trace over qubit $j$, we have
\[
\tr_j \ket{\psi'} \bra{\psi'}
 = q_j^0 \ket{0';(j)} \bra{0';(j)} + q_j^1 \ket{1';(j)} \bra{1';(j)} .
\]
Since this must be equal to $\rho_{(j)}$, it must have the eigenvalues
of $\rho_{(j)}$,
$q_j^0 = p_j^0$ and $q_j^1 = p_j^1$.
We consider two cases, depending on whether $\rho_{(j)}$ has
distinct eigenvalues or not.
Let us treat first the case of distinct eigenvalues,
$p_j^0 \neq p_j^1$.
In this case, the eigenvector $\ket{0';(j)}$
can be off by at most a phase from the eigenvector $\ket{0;(j)}$, and
similarly for $\ket{1';(j)}$.
The same argument applied to the one-qubit reduced density matrix
$\rho_j$ shows that $\ket{0'}$ can be off by at most a phase
from $\ket{0}$, and similarly for $\ket{1'}$.
In this case, then, we can write
\begin{equation}
\label{psiprime}
\ket{\psi'} = \sqrt{p_j^0} (L_j \ket{0}) \otimes_j \ket{0;(j)}
            + \sqrt{p_j^1} (L_j \ket{1}) \otimes_j \ket{1;(j)} ,
\end{equation}
with $L_j$ a $2 \times 2$ diagonal unitary matrix.

Let us treat next the case of repeated eigenvalues,
$p_j^0 = p_j^1$.
In this case,
the eigenvectors $\ket{0';(j)}$ and $\ket{1';(j)}$
must merely span the same two-dimensional complex space
that is spanned by $\ket{0;(j)}$ and $\ket{1;(j)}$.
In this case, the primed eigenvectors must be related to
the unprimed eigenvectors by a two-dimensional unitary transformation,
\begin{align*}
\ket{0';(j)} &= u_{00} \ket{0;(j)} + u_{01} \ket{1;(j)} \\
\ket{1';(j)} &= u_{10} \ket{0;(j)} + u_{11} \ket{1;(j)}
\end{align*}
with
\[
\twotwo{u_{00}}{u_{01}}{u_{10}}{u_{11}} \in U(2) .
\]
The same argument applied to the one-qubit reduced density matrix
$\rho_j$ shows that there must be some $2 \times 2$ unitary matrix
$v_{lm}$ with
\begin{align*}
\ket{0'} &= v_{00} \ket{0} + v_{01} \ket{1} \\
\ket{1'} &= v_{10} \ket{0} + v_{11} \ket{1} .
\end{align*}
In this case, then, we can write equation (\ref{psiprime})
with $L_j$ the $2 \times 2$ unitary matrix equal to the
product of the transpose of $v_{lm}$ with $u_{lm}$.
(We have abused notation
by using the symbol $L_j$ to represent
both the $2 \times 2$ unitary matrix, and also
the local unitary transformation on $n$-qubit state vectors
\[
I \otimes \cdots \otimes I \otimes L_j \otimes I \otimes \cdots
                               \otimes I ,
\]
with the $2 \times 2$ matrix $L_j$ in the $j$th slot of this
tensor product, and one-qubit ($2 \times 2$)
identity operators in all other slots.)
This completes the proof of the Claim.

For each pair of qubit labels $j,k$, we have
\[
\ket{\psi} = L_k^{-1} L_j \ket{\psi} .
\]
Next, spectrally decompose each $L_j$ with unitary
matrices $U_j$ so that
\[
D_j = U_j L_j U_j^{-1}
\]
are diagonal.
We have
\begin{align*}
D_k^{-1} D_j U_1 \cdots U_n \ket{\psi}
 &= U_1 \cdots U_n U_k^{-1} D_k^{-1} U_k U_j^{-1} D_j U_j \ket{\psi} \\
 &= U_1 \cdots U_n \ket{\psi} .
\end{align*}
Using the multi-index $I = (i_1 i_2 \cdots i_n)$, where each
$i_j$ is zero or one, and the basis
\[
\ket{I} = \ket{i_1 i_2 \cdots i_n}
 = \ket{i_1} \otimes \ket{i_2} \otimes \cdots \otimes \ket{i_n} ,
\]
expand
\[
U_1 \cdots U_n \ket{\psi} = \sum_I c_I \ket{I}
\]
and write
\[
D_j = e^{i \alpha_j} \twotwo{e^{i \beta_j}}{0}{0}{e^{-i \beta_j}} ,
\]
where $e^{i \beta_j} \neq e^{-i \beta_j}$, since $\ket{\psi'}$ is
a different state from $\ket{\psi}$.
Now we have
\[
c_I = c_I \exp\{i [\alpha_j - \alpha_k
             + (-1)^{i_j} \beta_j - (-1)^{i_k} \beta_k]\}
\]
for all multi-indices $I$ and all $j,k \in \{1,\ldots,n\}$.
We see that for each multi-index $I$, either $c_I = 0$
or
\[
\exp\{i [\alpha_j - \alpha_k + (-1)^{i_j} \beta_j - (-1)^{i_k} \beta_k]\} = 1
\]
for all $j,k \in \{1,\ldots,n\}$.
Let $J$ be a multi-index with $c_J \neq 0$.  If $I$ is any multi-index
that agrees with $J$ in at least one entry
(say the $j$th qubit entry), and disagrees with $J$
in at least one entry (say the $k$th qubit entry),
then $c_I = 0$.
We conclude that $c_J$ and $c_{\overline{J}}$,
where $\overline{J}$ is the multi-index consisting of
the complements of each of the $n$ bits in multi-index $J$,
are the only nonzero
coefficients in $U_1 \cdots U_n \ket{\psi}$.
Consequently,
$\ket{\psi}$ is LU-equivalent to a generalized $n$-qubit GHZ state.
\end{proof}

Much of this argument carries over to the more general situation
of a system of $n$ parties in which party $j$ has dimension
$d_j$ (party $j$ is a qubit if and only if $d_j = 2$).
In particular, the Claim carries over.
Suppose that $\ket{\psi}$ and $\ket{\psi'}$
are states of a system of $n$ parties in which party $j$ has
dimension $d_j$.
If $\ket{\psi}$ and $\ket{\psi'}$ have the same reduced
density matrices, then for each party $j \in \{1,\ldots,n\}$,
there is a one-party local unitary transformation $L_j$ such that
$\ket{\psi'} = L_j \ket{\psi}$.
The second part of the argument is complicated by the possibility
of repeated eigenvalues in the transformations $D_j$.
We leave this as a question for future work.

It is worthwhile to point out the significance of
stabilizer subgroups of the local unitary group in this work.
The local unitary group for $n$-qubit density matrices is the group
$G = SU(2)^n$, consisting of a special unitary
transformation on each qubit.
Each (pure or mixed) state $\rho$ has a stabilizer subgroup
$I_\rho$ consisting of elements of $G$ that leave $\rho$
fixed under the action $g \rho g^{-1}$ for $g \in G$.
We have seen that a state that is undetermined among pure states
has a special type of enlarged stabilizer subgroup.

There are two alternative formulations of the main result
that may suggest promising avenues for the classification
of entanglement types.
One can precisely characterize the pure $n$-qubit states
that are undetermined among pure states
in terms of the
stabilizer subalgebra of the state, that is the Lie algebra
of the stabilizer subgroup of the state.
It is shown in \cite{walcklyons07} that
the generalized $n$-qubit GHZ state
has (for $n \geq 3$) stabilizer subalgebra
\[
K_\rho = \left\{ \sum_{j=1}^n i t_j Z_j \left| \sum_{j=1}^n t_j = 0
          \right. \right\} ,
\]
where $Z_j$ is the Pauli matrix $\sigma_z$ applied to qubit $j$.
States of $n$ qubits undetermined among pure states are precisely
those which are LU-equivalent to states with this subalgebra.

A second alternative formulation of the main result is in terms
of the dimension of the stabilizer subgroup.
For $n = 3$ and $n \geq 5$, an $n$-qubit state
is undetermined among pure states
if and only if it is not a product state
and its stabilizer subgroup has dimension $n-1$
\cite{lwb}.

We have shown that all $n$-qubit
states other than generalized $n$-qubit GHZ states and their LU-equivalents
are completely determined by their reduced density matrices.
Is it necessary to specify \emph{all} of the reduced density matrices?
Which states are undetermined by specifying only
$n-1$ (rather than all $n$)
of the $(n-1)$-qubit reduced density matrices?
Is it a larger set than the generalized GHZs?
The answer is yes, and
stabilizers can help us understand this.
For example, the state
\[
\ket{\chi} = \frac{1}{\sqrt{3}} (\ket{0000} + \ket{0001} + \ket{1111})
\]
is undetermined by its 3-qubit reduced density matrices
obtained by taking the partial trace over qubit 1,
the partial trace over qubit 2, and
the partial trace over qubit 3.
It is not LU-equivalent to a generalized 4-qubit GHZ state
(it has a different stabilizer subalgebra structure, and stabilizer
subalgebra structure is an LU-invariant).
Note that $Z_1 \ket{\chi} = Z_2 \ket{\chi} = Z_3 \ket{\chi}
 \neq e^{i \alpha} \ket{\chi}$.
The state $Z_1 \ket{\chi}$ has the same 3-qubit reduced density
matrices as $\ket{\chi}$ when taking the partial trace over
qubit 1, 2, or 3.  These two states have a different 3-qubit
reduced density matrix when taking the partial trace over qubit 4.

A pure state's LU-equivalence class is often considered to contain all
of the information about the entanglement of the state.  An
interesting question is: Are there $n$-qubit pure states with
entanglement information that is not contained in their reduced
density matrices?  We might interpret this question as equivalent to
the question: Are there $n$-qubit pure states for which the
LU-equivalence class of the state is undetermined by its reduced
density matrices?  The answer to this question is no.  Every $n$-qubit
pure state can be determined (among pure states) up to a local unitary
transformation by its reduced density matrices.  This can be seen
directly from the Claim at the beginning of our proof.  Two pure
states that have the same reduced density matrices must be LU
equivalent.

We have not answered the question of which $n$-qubit pure states
are undetermined among arbitrary states by their reduced
density matrices.
The set of $n$-qubit states undetermined
among arbitrary states must contain the generalized GHZ states,
bit it could be strictly larger than
the set that is undetermined among pure
states.  This remains an open question.

The authors thank the National Science Foundation
for their support of this work through NSF Award No. PHY-0555506.



\end{document}